\documentclass[runningheads]{llncs}
\usepackage[T1]{fontenc}
\usepackage{graphicx}
\usepackage{utfsym}
\usepackage{multirow}
\usepackage{colortbl} 
\definecolor{mygray}{gray}{0.9}
\definecolor{newcolor}{rgb}{.8,.349,.1}
\usepackage{marvosym}
\usepackage{amsmath}
\usepackage{amssymb}
\usepackage[hyperfootnotes=false]{hyperref}

\begin{document}
\title{Advancing UWF-SLO Vessel Segmentation with Source-Free Active Domain Adaptation and a Novel Multi-Center Dataset}

%

\author{Hongqiu Wang\inst{1}* \and
Xiangde Luo\inst{2}* \and
Wu Chen\inst{2} \and
Qingqing Tang\inst{3}\and
Mei Xin\inst{4}\and
Qiong Wang \inst{5} \and
Lei Zhu\inst{1,6}\textsuperscript{(\Letter)}}
\authorrunning{H. Wang et al.}
%
\institute{Hong Kong University of Science and Technology (Guangzhou), Guangzhou, China \email{hwang007@connect.hkust-gz.edu.cn} \and
 University of Electronic Science and Technology of China, Chengdu, China\\
\and
West China Hospital, Sichuan University, Chengdu, China\\
\and 
Chengdu First People's Hospital\\
\and
Shenzhen Institutes of Advanced Technology, Chinese Academy of Sciences\\
\and 
The Hong Kong University of Science and Technology\\
}

\maketitle 
\footnote{H. Wang and X. Luo contributed equally to this work.}
\vspace{-8mm}
\begin{abstract}
Accurate vessel segmentation in Ultra-Wide-Field Scanning Laser Ophthalmoscopy (UWF-SLO) images is crucial for diagnosing retinal diseases. Although recent techniques have shown encouraging outcomes in vessel segmentation, models trained on one medical dataset often underperform on others due to domain shifts. Meanwhile, manually labeling high-resolution UWF-SLO images is an extremely challenging, time-consuming and expensive task. In response, this study introduces a pioneering framework that leverages a patch-based active domain adaptation approach. By actively recommending a few valuable image patches by the devised Cascade Uncertainty-Predominance (CUP) selection strategy for labeling and model-finetuning, our method significantly improves the accuracy of UWF-SLO vessel segmentation across diverse medical centers. In addition, we annotate and construct the first Multi-center UWF-SLO Vessel Segmentation (MU-VS) dataset to promote this topic research, comprising data from multiple institutions. This dataset serves as a valuable resource for cross-center evaluation, verifying the effectiveness and robustness of our approach. Experimental results demonstrate that our approach surpasses existing domain adaptation and active learning methods, considerably reducing the gap between the Upper and Lower bounds with minimal annotations, highlighting our method's practical clinical value. We will release our dataset and code to facilitate relevant research (\href{https://github.com/whq-xxh/SFADA-UWF-SLO}{\textit{Git}}).
\end{abstract}
\section{Introduction}
Accurate segmentation of retinal vessels in fundus images is critical in aiding ophthalmologists with quantitative analysis and treatment \cite{fraz2012blood,mookiah2021review}. For instance, Retinal Vein Occlusion (RVO) is identified by increased retinal vessel tortuosity, enlarged vessel caliber, and retinal non-perfusion \cite{rogers2010prevalence}. A variety of deep learning models for automated vessel segmentation have emerged, showing promising results \cite{galdran2018no,menten2022physiology,xing2023diff,xing2024segmamba,xu2020boosting}. These segmentation models are mainly adapted to Narrow Field (NF) Fundus Photography (FP), since NF FP is the most common format and modality in clinical practice, and previous relevant segmentation datasets are also mainly based on NF FP \cite{hoover2000locating,staal2004ridge,zhang2016robust}.

Recently, Ultra-Wide-Field Scanning Laser Ophthalmoscopy (UWF-SLO) imaging has gained popularity due to its ability to provide extensive retinal coverage and superior imaging of peripheral lesions over NF FP, thereby enhancing diagnostic precision \cite{tang2024applications}. Generally, UWF-SLO images provide an expansive 200° field-of-view (FOV), far exceeding the typical 30°-50° FOV of NF FP. This broader view grants ophthalmologists access to more comprehensive information for more accurate diagnoses \cite{ding2020weakly,nagiel2016ultra}. There are also some efforts in the field. For example, Li \textit{et al.} proposed a weakly-supervised iterative learning method and the PRIME-P20 dataset to segment vessels in UWF images\cite{ding2020weakly}. Qiu \textit{et al.} introduced a dual-stream super-resolution network for this task\cite{qiu2023rethinking}.

\begin{table}[t]
\centering
\caption{Quantitative analysis of the existing datasets, including categories and image resolution. Center A: Hospital A. Center B: Hospital B.}
\vspace{-3mm}
\label{tab_data}
\resizebox{0.9\textwidth}{!}{%
\begin{tabular}{c|c|c|c|c }
\hline
Dataset & Amount of data & Categories & Resolution & Public available \\
\hline
In-house \cite{li2023privileged} & 65 & Normal,VO & 3900$\times$3072 & no \\
PRIME-FP20 \cite{ding2020weakly} & 15 & DR & 4000$\times$4000 & yes \\
\textbf{Center A (Ours) }& 30 & Normal, RVO & 3900$\times$3072 & yes \\
\textbf{Center B (Ours)} & 30 & DR, RVO, RP, RAO, CSC& 3900$\times$3072  &  yes \\
\hline
\end{tabular}}
\vspace{-3mm}
\end{table}

\begin{figure}[t]
    \centering
    \includegraphics[width=1\textwidth]{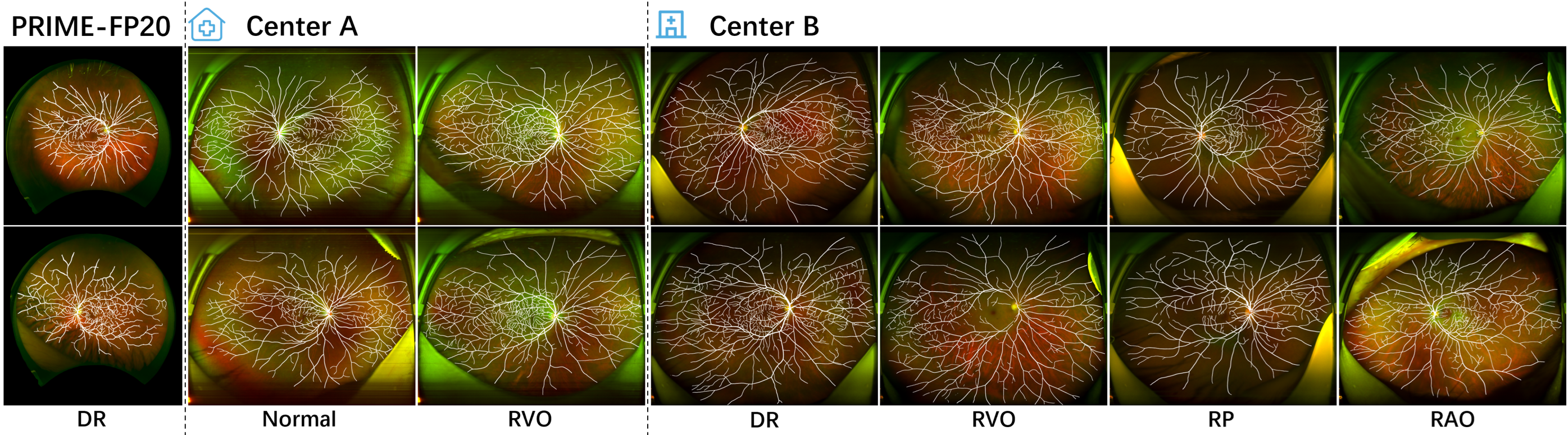}
    \vspace{-7mm}
    \caption{Several data visualization examples from the proposed Multi-center UWF-SLO Vessel Segmentation (MU-VS) dataset, illustrate that various centers encompass different disease categories.}
    \vspace{-6mm}
    \label{fig:F1}
\end{figure}

Although there has been a notable improvement in UWF vessel segmentation approaches, the above models are all developed within the single-center setting, lacking multi-center, cross-domain studies. In real-world clinical scenarios, domain shifts commonly occur among datasets \cite{guan2021domain,zhao2024morestyle} because of variations in imaging equipment and patient populations, potentially leading to suboptimal performance in new domains \cite{niu2024survey,wang2023advancing}. The simplest approach involves fully supervised training on target data, yet annotating high-resolution UWF-SLO images is extremely expensive, requiring approximately 18 hours of an expert's time to annotate a single image \cite{pellegrini2014blood}.
Therefore, unsupervised domain adaptation (UDA) techniques are widely explored, designed to reduce the domain discrepancy between the labeled source and the unlabeled target domain \cite{kumari2023deep}. Although UDA approaches yield better outcomes, their performance still significantly falls short of that achieved by fully supervised models \cite{kumari2023deep,wang2023advancing}. Moreover, accessing source medical datasets raises privacy and security concerns \cite{zhang2023source}.

To alleviate these above issues, we propose a novel patch-based Source-Free Active Domain Adaptation (SFADA) method for advancing UWF-SLO cross-center vessel segmentation. Our approach offers three advantages: First, it eliminates the need to access source domain data, thereby enhancing data security and privacy protection. Second, we introduce the Cascade Uncertainty-Predominance (CUP) selection strategy, which efficiently identifies a small subset of valuable image patches for annotation, substantially reducing the annotation burden. Lastly, by integrating our method with a minimal number of patch annotations, we can significantly boost the model's performance. Meanwhile, by integrating the existing dataset and our newly collected and labeled datasets from two distinct centers (as detailed in Table~\ref{tab_data} and Fig.~\ref{fig:F1}), we construct the first Multi-center UWF-SLO Vessel Segmentation (MU-VS) dataset to explore the cross-center segmentation study. The main contributions are summarized as follows:
\vspace{-5.5mm}
\begin{itemize}
\item To our knowledge, this marks a pioneering exploration on an essential application of cross-center vessel segmentation using UWF-SLO, and we propose a patch-based SFADA framework to enhance segmentation performance.
\item We design the Cascade Uncertainty Predominance (CUP) selection strategy to select a small number of patches with high uncertainty and dominance to recommend for manual annotation.
\item We establish the first multi-center UWF-SLO vessel segmentation dataset consisting of 60 UWF-SLO images from two hospitals, named MU-VS, to support relevant studies.
\item Experimental results show that our method significantly surpasses other state-of-the-art domain adaptation and active learning methods, effectively enhancing segmentation accuracy.
\end{itemize}

\section{Methodology}
\subsection{Problem setting}
The goal of medical image segmentation is to construct a model $\mathcal{M}$ that links an image sample $x$ from the space $X$ to its predictive label $y$ within the space $Y$. In the SFADA setting, direct access to the source dataset and its annotations ${(x^s, y^s)}$ is avoided, thereby safeguarding data privacy and security. Instead, we employ a model $\mathcal{M}^s$ pre-trained in the source domain alongside unlabeled data $X^t$ from the target domain to guide the recommendation of annotations.
The quantity of target patches chosen for manual annotation is denoted by $N_t^{AL} = \alpha \cdot N_t$, with $N_t^{AL} \ll N_t $ indicating that the actively selected patches are significantly fewer than the total target patches, where $\alpha$ signifies the selection ratio and $N_t$ is the count of all target patches. Concurrently, the labels for these selected patches are symbolized as $Y^{Lt}$. Our goal is to refine the performance of the model $\mathcal{M}^t$ in the target domain, striving to keep the parameter $\alpha$ as small as possible.

\begin{figure}[t]
    \centering
    \includegraphics[width=1\textwidth]{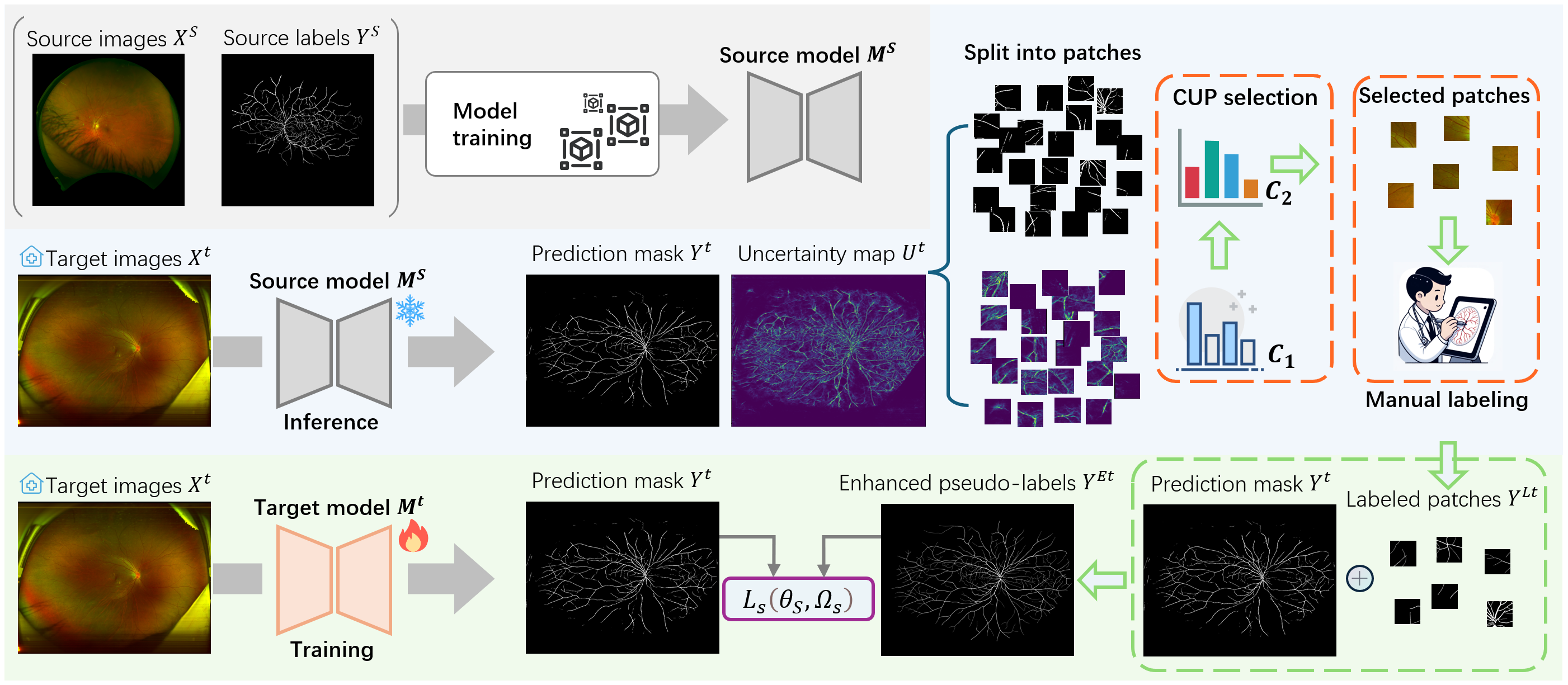}
    \vspace{-8mm}
    \caption{Overview of the proposed patch-based SFADA framework including the CUP selection strategy. The first row of gray represents the training of source model; the second row of light blue represents the recommendation and annotation of the valuable image patches based on the CUP strategy; and the third row of green represents the target model's fine-tuning under the supervision of enhanced pseudo-labels.}
    \label{fig:Overview}
    \vspace{-4mm}
\end{figure}

\subsection{Pipeline of Patch-Based SFADA Framework}
Considering the high-resolution UWF-SLO images used in our segmentation task, fully annotating the image is extremely expensive and time-consuming. To address this, we propose a method that focuses on selecting and annotating a few valuable image patches and finally integrating them into enhanced pseudo-labels ($Y^{Et}$) for target model $\mathcal{M}^t$ fine-tuning.

Fig.~\ref{fig:Overview} depicts the structural design of our patch-based SFADA framework. The initial row illustrates the pipeline's first phase, where we utilize images with their corresponding annotations ${(X^s, Y^s)}$ from the source domain to train the segmentation model, obtaining the source domain segmentation model $\mathcal{M}^s$. The second step is depicted in the second row of Fig.~\ref{fig:Overview}. Here, we freeze the parameters of $\mathcal{M}^s$ to infer the target domain UWF-SLO image $X^t$, subsequently deriving its prediction masks $Y^t$ and uncertainty maps $U^t$ (detailed computational methods are explained in Section~\ref{cup}). Subsequently, we divide $Y^t$ and $U^t$ into numerous small patches and recommend the most valuable ones to ophthalmologists for annotation, employing our CUP strategy.
The third step, illustrated in the third row of Fig.~\ref{fig:Overview}, involves merging the actively annotated real patch labels $Y^{Lt}$ with the network's prediction masks $Y^t$ to produce the enhanced pseudo labels $Y^{Et}$. The specific operation is to modify the corresponding image patches using $Y^{Lt}$ according to the position index of these patches. Finally, the source domain model $\mathcal{M}^s$ undergoes fine-tuning guided by the enhanced pseudo-label $Y^{Et}$ to develop the target domain model $\mathcal{M}^t$. This process involves minimizing the loss function $L_s(\theta_s, \Omega_s)$ (a combined loss function of cross-entropy and Dice) in relation to the network parameters $\Omega_s$.

\subsection{Cascade Uncertainty-Predominance selection}\label{cup}
In the domain adaptation task, the source domain model $\mathcal{M}^s$ has already acquired some fundamental knowledge of vessel segmentation, enabling it to generate a preliminary vessel mask from target domain data $X^{t}$. However, the variances between domains result in several regions within the $\mathcal{M}^s$ predicted vessel mask exhibiting high uncertainty. To address this challenge, we design a CUP selection strategy to prioritize patches with higher uncertainty, where $\mathcal{M}^s$ may lack related knowledge. Building on this premise, we further identify patches with substantial vessel prediction by $\mathcal{M}^s$, denoting regions of predominance. This approach underscores that, despite significant uncertainties in $\mathcal{M}^s$'s predictions, where model predictions are prone to errors. Certain regions still predict a large number of target vessels, necessitating ground truth annotations for precise model training and refinement.

As illustrated in the CUP selection box of Fig.~\ref{fig:Overview}, the CUP strategy comprises two cascades: $C_1$ for uncertainty and $C_2$ for predominance. First, we employ the source domain model $\mathcal{M}^s$ to generate the prediction masks $Y^t$ and the uncertainty maps $U^t$. The procedure for prediction masks is detailed as follows:
\begin{equation}
Prediction\_mask = \underset{c}{\arg\max} \left( \underset{c}{softmax}(\mathcal{M}^s(x^t)) \right),
\end{equation}
where $x^t$ represents a target image from $X^t$, and $c$ denotes the predicted category, here refers to the foreground vessels. The process for uncertainty maps is outlined in the following manner:
\begin{equation}
Uncertainty = -1 \times \sum_{c=1}^{C} \underset{c}{softmax}(\mathcal{M}^s(x^t)) \log \left( \underset{c}{softmax}(\mathcal{M}^s(x^t)) \right),
\end{equation}
where $\sum_{c=1}^{C}$ represents the sum of $c$ classes, both the foreground vessels and background. This measure reflects the entropy of the model's predictive probability distribution: higher values indicate greater uncertainty (i.e., the model's predictions are spread out across different classes).

Next, as depicted in Fig.~\ref{fig:Overview}, the prediction masks $Y^t$ and the uncertainty maps $U^t$ are divided into multiple small patches. For each patch, we calculate the total number of predicted vessel pixels and the aggregate uncertainty values, denoted as $ves_{p}=\{P_{0},..., P_t\}$ and $ves_{u}=\{U_{0},..., U_t\}$, respectively. Based on these statistical results, our cascade selection strategy is operated as follows:

\begin{equation}
    \begin{aligned}
    &\text{\hspace*{0pt}step1: }Select_{C1\%} = \text{Top}(\text{ves}_u)[C1\%], \\
    &\text{\hspace*{0pt}step2: }Selected = \text{Top}(\text{ves}_p\ in\ Select_{C1\%})[C2\%],
    \end{aligned}
\end{equation}where $C1\%$ and $C2\%$ denote the ratio of patches selected based on the highest uncertainty and predicted vessel pixels, respectively.

\section{Experiments and Results}
\subsection{Data Description}
We collected 30 UWF-SLO images each from two distinct medical centers, utilizing Optos California and 200Tx cameras (Optos plc, Dunfermline, UK) for capture. The datasets from each center comprised varied categories (refer to Table~\ref{tab_data} for details) and were annotated with vessel masks by their respective ophthalmologists. The ophthalmologists utilized Photoshop software for the manual annotation of vessels within the UWF-SLO images. They precisely labeled the vessels across different regions by iteratively fine-tuning the image's brightness and contrast, adopting a layered approach, and leveraging the software's outlining tools for accurate delineation. Combined with the currently existing publicly available data PRIME-FP20 \cite{ding2020weakly}, we established the first multi-center vessel segmentation of UWF-SLO, with domain shifts potentially attributed to different annotators and different annotation approaches and disease categories.

\subsection{Implementation Details and Evaluation Metrics}
\noindent\textbf{Implementation Details.}
For objective evaluation, each dataset is randomly split into three subsets (training, validation, testing) with a ratio of 6:2:2. The model that performs best on the validation set is then chosen for reporting its results on the test set. The PRIME-FP20 \cite{ding2020weakly} dataset is used as the source domain, with centers A and B serving as the target domains. All experiments are carried out on an NVIDIA RTX 3090 GPU with 24 GB memory. The original image size is 3900$\times$3072 and the patch size is set to 260$\times$256. $C1\%$ and $C2\%$ are set to 10\% and 50\% respectively, which means that a total of $\alpha=5\%$ of the patches are selected for annotation. All input images resized to 1024$\times$1024 for uniform training. The SGD optimizer and a batch size of 5 are employed for training. For original training with all labels, models undergo 6000 iterations, while fine-tuning with pseudo labels involves 3000 iterations. An initial learning rate of 0.03 is set, undergoing exponential decay at a rate of 0.9 per iteration. For consistency, comparison methods are re-implemented using the same U-Net backbone \cite{ronneberger2015u} and executed under identical conditions.

\noindent\textbf{Evaluation Metrics.}
Following previous work \cite{qiu2023rethinking}, We employ the Dice score (Dice), Intersection over Union (IoU), Matthews Correlation Coefficient (MCC), and Bookmaker Informedness (BM) as metrics to assess the performance of these models. Higher values indicate superior model performance.

\begin{table*}[t]
\centering
\caption{Quantitative comparison on Dice and IoU of our method and other state-of-the-art domain adaptation and active learning methods on the MU-VS dataset.}
\vspace{-3mm}
\label{tab2}
\setlength\tabcolsep{3pt}
\resizebox{0.98\textwidth}{!}{%
\begin{tabular}{c|c c |c|c c |c}
\hline
\rowcolor{mygray} 
 & \multicolumn{3}{c|}{Dice (mean$\pm$std, \%)} & \multicolumn{3}{c}{IoU (mean$\pm$std, \%)}  \\
\hline
\hline
Methods & Center A & Center B & Overall & Center A & Center B & Overall \\
\hline
\hline
Lower bound & $54.76\pm2.73$ & $51.32\pm3.25$ & $53.04\pm2.99$ & $37.75\pm2.56$ & $34.58\pm2.91$ & $36.17\pm2.74$ \\
Upper bound & $62.00\pm1.32$ & $57.36\pm2.99$ & $59.68\pm2.16$ & $44.94\pm1.38$ & $40.27\pm2.96$ & $42.61\pm2.17$ \\
\hline
\hline
AdvEnt \cite{vu2019advent}& $56.29\pm1.35$ & $51.95\pm2.96$ & $54.12\pm2.16$ & $39.18\pm1.30$ & $35.14\pm2.72$ & $37.16\pm2.01$   \\
DPL \cite{chen2021source}& $56.50\pm2.20$ & $52.58\pm2.92$ & $54.54\pm2.56$ & $39.40\pm2.12$ & $35.72\pm2.67$ & $37.56\pm2.40$  \\
CBMT \cite{tang2023source}& $57.51\pm1.42$ & $52.95\pm2.79$ & $55.23\pm2.10$ & $40.38\pm1.39$ & $36.06\pm2.57$ & $38.22\pm1.98$ \\
CPR \cite{huai2023context}& $57.79\pm2.01$ & $53.28\pm2.93$ & $55.54\pm2.47$ & $40.66\pm1.96$ & $36.36\pm2.71$ & $38.51\pm2.34$  \\
\hline
\hline
Adversarial \cite{tsai2018learning}& $58.79\pm1.73$ & $53.32\pm2.76$ & $56.06\pm2.25$ & $41.66\pm1.71$ & $36.40\pm2.56$ & $39.03\pm2.14$ \\
AADA \cite{su2020active}& $58.92\pm1.46$ & $53.38\pm2.84$ & $56.15\pm2.15$ & $41.78\pm1.46$ & $36.45\pm2.66$ & $39.12\pm2.06$ \\
MHPL \cite{wang2023mhpl}& $59.32\pm1.22$ & $53.58\pm2.99$ & $56.45\pm2.11$ & $42.18\pm1.23$ & $36.66\pm2.84$ & $39.42\pm2.04$ \\
STDR \cite{wang2024dual} & $59.51\pm1.54$ & $53.96\pm2.80$ & $56.73\pm2.17$ & $42.38\pm1.56$ & $37.00\pm2.67$ & $39.69\pm2.11$  \\
\hline
\hline
Ours & $60.92\pm0.94$ & $54.92\pm2.81$ & $57.92\pm1.88$ & $43.81\pm0.98$ & $37.91\pm2.70$ & $40.86\pm1.84$ \\
\hline
\end{tabular}}
\vspace{-3mm}
\end{table*}

\subsection{Experimental Results}
\begin{figure}[h!]
    \centering
    \includegraphics[width=1\textwidth]{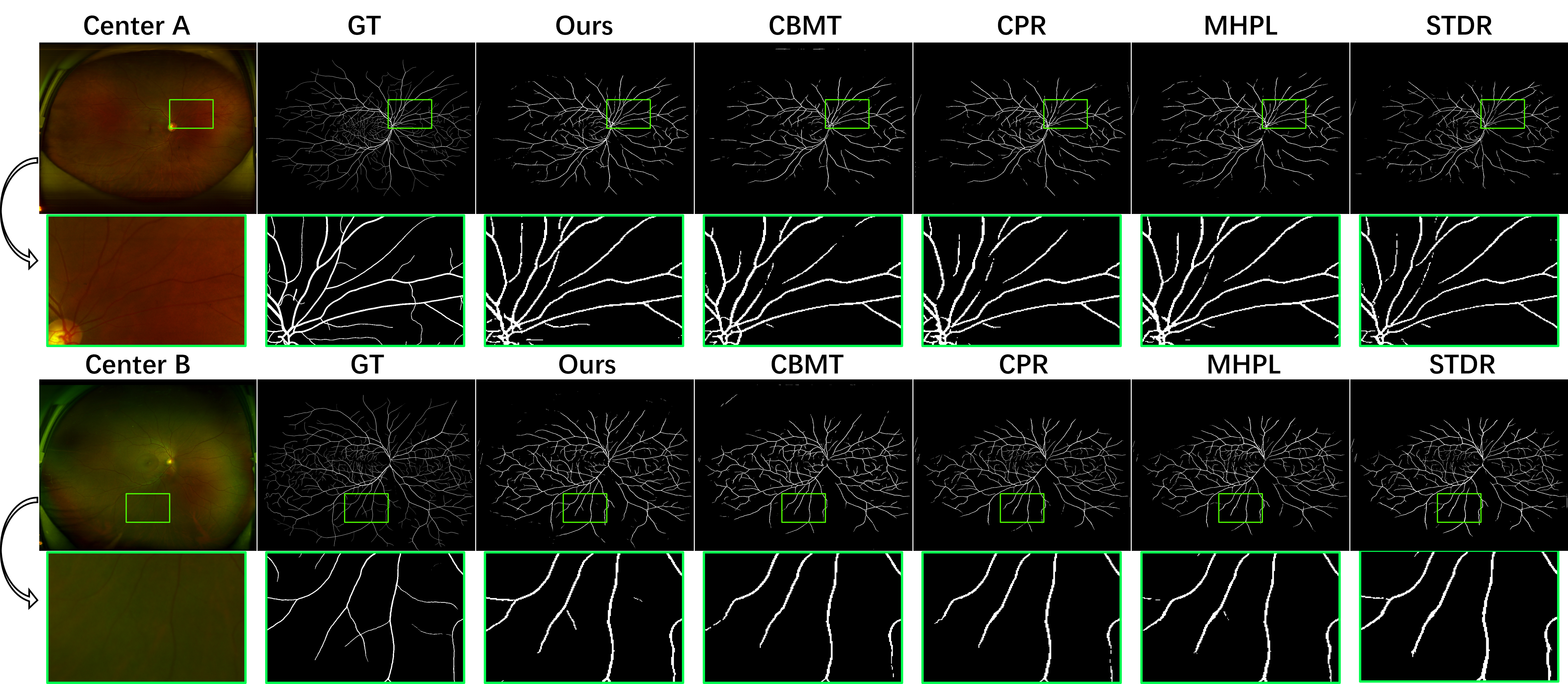}
    \vspace{-5mm}
    \caption{Visual comparisons of our method and other leading domain adaptation and active learning methods. Our method can more accurately segment the vessels (please zoom in for more details).}
    \vspace{-3mm}
    \label{fig:demo3}
\end{figure}

This section provides an overview of experimental results across various medical centers in Table~\ref{tab2} and~\ref{tab3}, including the lower bound (model without finetuning), upper bound (model finetuned with all labels), and comparisons with other state-of-the-art domain adaptation and active learning methods. Fig.~\ref{fig:demo3} exhibits some visualizations of the segmentation results. By analyzing the data from Table~\ref{tab2} and Table~\ref{tab3}, it becomes evident that significant performance gaps exist between the lower and upper bounds across various evaluation metrics. For instance, in the case of the BM metric, the overall gap widens from 49.61\% to 58.29\%.

\begin{table*}[t]
\centering
\caption{Quantitative comparison on MCC and BM of our method and other state-of-the-art domain adaptation and active learning methods on the MU-VS dataset.}
\vspace{-3mm}
\label{tab3}
\setlength\tabcolsep{3pt}
\resizebox{0.98\textwidth}{!}{%
\begin{tabular}{c|c c |c|c c |c}
\hline
\rowcolor{mygray} 
 & \multicolumn{3}{c|}{MCC (mean$\pm$std, \%)} & \multicolumn{3}{c}{BM (mean$\pm$std, \%)}  \\
\hline
\hline
Methods & Center A & Center B & Overall & Center A & Center B & Overall \\
\hline
\hline
Lower bound & $54.03\pm2.54$ & $50.33\pm3.25$ & $52.18\pm2.90$ & $48.95\pm4.91$ & $50.27\pm4.74$ & $49.61\pm4.83$ \\
Upper bound & $61.48\pm1.29 $ & $56.58\pm2.80$ & $59.03\pm2.05$ & $55.71\pm3.98$ & $60.87\pm4.23$ & $58.29\pm4.10$ \\
\hline
\hline
AdvEnt \cite{vu2019advent}& $55.40\pm1.38$ & $50.94\pm2.85$ & $53.17\pm2.11$ & $51.18\pm3.63$ & $52.92\pm4.64$ & $52.05\pm4.14$  \\
DPL \cite{chen2021source}& $55.69\pm2.13$ & $51.60\pm2.79$ & $53.65\pm2.46$ & $51.07\pm4.39$ & $51.39\pm4.59$ & $51.23\pm4.49$ \\
CBMT \cite{tang2023source}& $56.79\pm1.42$ & $51.99\pm2.67$ & $54.39\pm2.05$ & $51.25\pm2.84$ & $52.34\pm4.68$ & $51.79\pm3.76$  \\
CPR \cite{huai2023context}& $57.21\pm1.93$ & $52.35\pm2.82$ & $54.78\pm2.38$ & $50.88\pm3.67$ & $55.33\pm4.73$ & $53.11\pm4.20$ \\
\hline
\hline
Adversarial \cite{tsai2018learning}& $57.99\pm1.79$ & $52.39\pm2.64$ & $55.19\pm2.22$ & $53.27\pm3.27$ & $55.60\pm4.31$ & $54.44\pm3.79$ \\
AADA \cite{su2020active}& $58.03\pm1.54$ & $52.43\pm2.70$ & $55.23\pm2.12$ & $53.99\pm3.00$ & $54.63\pm4.88$ & $54.31\pm3.94$ \\
MHPL \cite{wang2023mhpl}& $58.38\pm1.31$ & $52.72\pm2.89$ & $55.55\pm2.10$ & $55.00\pm2.74$ & $56.91\pm5.00$ & $55.96\pm3.87$ \\
STDR \cite{wang2024dual}& $58.58\pm1.64$ & $53.07\pm2.67$ & $55.83\pm2.16$ & $55.08\pm2.97$ & $56.56\pm4.78$ & $55.82\pm3.88$  \\
\hline
\hline
Ours & $59.94\pm0.99$ & $54.09\pm2.67$ & $57.02\pm1.83$ & $57.82\pm3.06$ & $57.95\pm5.15$ & $57.89\pm4.10$ \\
\hline
\end{tabular}}
\vspace{-3mm}
\end{table*}

\begin{table*}[t]
\centering
\caption{Ablation experiments on the MU-VS dataset with Dice and IoU.}
\vspace{-2mm}
\label{tab4}
\setlength\tabcolsep{3pt}
\resizebox{1\textwidth}{!}{%
\begin{tabular}{c|c c c|c c |c|c c |c}
\hline
\rowcolor{mygray} 
& \multicolumn{3}{c|}{Patch-based Methods} & \multicolumn{3}{c|}{Dice (mean$\pm$std, \%)} & \multicolumn{3}{c}{IoU (mean$\pm$std, \%)}  \\
\hline
\hline
Methods & Random & C1 & C2 & Center A & Center B & Overall & Center A & Center B & Overall \\
\hline
\hline
M1 & $\checkmark$ & - & - & $58.17\pm1.89$ & $52.87\pm2.78$ & $55.52\pm2.34$ & $41.04\pm1.86$ & $35.99\pm2.56$ & $38.52\pm2.21$  \\
M2& - & $\checkmark$ & - & $59.13\pm1.42$ & $53.52\pm2.82$ & $56.33\pm2.12$ & $41.99\pm1.43$ & $36.58\pm2.62$ & $39.29\pm2.03$ \\
Ours& - & $\checkmark$ & $\checkmark$ & $60.92\pm0.94$ & $54.92\pm2.81$ & $57.92\pm1.88$ & $43.81\pm0.98$ & $37.91\pm2.70$ & $40.86\pm1.84$ \\
\hline
\end{tabular}}
\vspace{-3mm}
\end{table*}

\begin{table*}[t]
\centering
\caption{Ablation experiments on the MU-VS dataset with MCC and BM.}
\vspace{-2mm}
\label{tab5}
\setlength\tabcolsep{3pt}
\resizebox{1\textwidth}{!}{%
\begin{tabular}{c|c c c|c c |c|c c |c}
\hline
\rowcolor{mygray} 
& \multicolumn{3}{c|}{Patch-based Methods} & \multicolumn{3}{c|}{MCC (mean$\pm$std, \%)} & \multicolumn{3}{c}{BM (mean$\pm$std, \%)}  \\
\hline
\hline
Methods & Random & C1 & C2 & Center A & Center B & Overall & Center A & Center B & Overall \\
\hline
\hline
M1 & $\checkmark$ & - & - & $57.46\pm1.88$ & $51.93\pm2.66$ & $54.69\pm2.27$ & $51.97\pm3.51$ & $54.86\pm4.80$ & $53.42\pm4.15$  \\
M2& - & $\checkmark$ & - & $58.17\pm1.52$ & $52.68\pm2.66$ & $55.42\pm2.09$ & $54.93\pm3.03$ & $56.93\pm4.64$ & $55.93\pm3.83$ \\
Ours& - & $\checkmark$ & $\checkmark$ & $59.94\pm0.99$ & $54.09\pm2.67$ & $57.02\pm1.83$ & $57.82\pm3.06$ & $57.95\pm5.15$ & $57.89\pm4.10$\\
\hline
\end{tabular}}
\vspace{-5mm}
\end{table*}

\noindent\textbf{Domain adaptation analysis.} We benchmark our method against the latest state-of-the-art domain adaptation techniques under identical backbone architectures and experimental conditions. This comparison encompasses methods requiring access to source data, such as AdvEnt \cite{vu2019advent}, and source-free approaches like DPL \cite{chen2021source}, CBMT \cite{tang2023source}, and CPR \cite{huai2023context}. The experimental outcomes indicate that various methods have led to improvements in accuracy. As illustrated in Table~\ref{tab2}, the overall Dice scores for these techniques vary from 54.12\% to 55.54\%, surpassing the lower bound of 53.04\%. However, perhaps due to the absence of supervised training with real labels, these improvements are relatively limited, and our method achieved 57.92\% in this indicator with few labels (5\%).

\noindent\textbf{Active learning analysis.} Given that SFADA incorporates elements of active learning, we also compare our approach with the recent leading active learning methods, all evaluated under the same experimental setup with 5\% labeled data. This comparison includes methods such as Adversarial \cite{tsai2018learning}, AADA \cite{su2020active}, MHPL \cite{wang2023mhpl}, and STDR \cite{wang2024dual}, ensuring a comprehensive analysis under uniform conditions. Merging the data from Table~\ref{tab2} and Table~\ref{tab3} reveals that, overall, active learning approaches outperform domain adaptation methods. Notably, our strategy yields the highest scores across all four metrics, underscoring the efficacy of our patch-based approach augmented by the CUP selection strategy.

\noindent\textbf{Ablation Studies.} To verify the effectiveness of our method, we conduct corresponding ablation experiments (the results are shown in Table~\ref{tab4} and Table~\ref{tab5}), including three configurations as follows: (1) M1: randomly select patches combined with our patch-based framework. (2) M2: based on $C1$ uncertainty, the top 5\% patches are selected for annotation and then integrated into the overall framework. (3) Ours: performing cascade selection, first select $C1$ uncertainty and then $C2$ predominance, named the CUP strategy. The experimental results show that M2 outperforms M1 overall, while Ours achieves greater performance gains compared to M2, e.g., the overall Dice from 55.52\% to 56.33\%, and finally to 57.92\% in Table~\ref{tab4}, which proves the effectiveness of our CUP strategy.

\section{Conclusion}
In this paper, we explore the task of vessel segmentation of UWF-SLO images across different centers. Considering the high cost of labeling high-resolution UWF-SLO images, we propose a patch-based SFADA approach to significantly save labeling resources while boosting segmentation performance. We also devise a CUP strategy to cascade the selection of valuable patches with high uncertainty and dominance for annotation. Meanwhile, we construct the first public multi-center UWF-SLO vessel segmentation (MU-VS) dataset to facilitate related research. Experimental results demonstrate that our method achieves optimal results compared to other domain adaptation and active learning methods. In the future, we plan to extend our method to other similar high-resolution medical image segmentation tasks.

\bibliographystyle{splncs04}
\bibliography{ref}
%




\end{document}